\documentclass[runningheads]{llncs}
\usepackage{graphicx}
\usepackage{cite}
\usepackage{amsmath,amssymb,amsfonts}
\usepackage{algorithmic}
\usepackage{textcomp}
\usepackage{xcolor}
\usepackage{multirow}
\begin{document}
\title{
A Transformer-based Framework for POI-level Social Post Geolocation
}
\titlerunning{A Transformer-based Framework for POI-level Social Post Geolocation}
%
\author{
Menglin Li\inst{1} \and
Kwan Hui Lim\inst{1} \and
Teng Guo\inst{2}\and
Junhua Liu\inst{1,3}
}
\authorrunning{M. Li et al.}

\institute{
Singapore University of Technology and Design, Singapore \and 
Dalian University of Technology, China \and
Forth AI, Singapore \\
\email{menglin\_li@mymail.sutd.edu.sg, kwanhui\_lim@sutd.edu.sg, teng.guo@outlook.com, j@forth.ai}
}

\maketitle              
\begin{abstract}
POI-level geo-information of social posts is critical to many location-based applications and services. 
However, the multi-modality, complexity and diverse nature of social media data and their platforms limit the performance of inferring such fine-grained locations and their subsequent applications.  
To address this issue, we present a transformer-based general framework, which builds upon pre-trained language models and considers non-textual data, for social post geolocation at the POI level. 
To this end, inputs are categorized to handle different social data, and an optimal combination strategy is provided for feature representations. 
Moreover, a uniform representation of hierarchy is proposed to learn temporal information, and a concatenated version of encodings is employed to capture feature-wise positions better. 
Experimental results on various social datasets demonstrate that three variants of our proposed framework outperform multiple state-of-art baselines by a large margin in terms of accuracy and distance error metrics.

\keywords{Information Retrieval \and Social Media \and Transformer \and Geolocation.}
\end{abstract}

\section{Introduction}
Knowing the posting location of social media data is important for many useful applications, including local event/place recommendations\cite{liu2020strategic}, location-based advertisements, emergency location identification, and natural disaster response\cite{zheng2018survey,liu2021crisisbert}.
However, geotagged social posts are very limited as less than 1\% of tweets are labeled with geo-coordinates\cite{cheng2010you}. This constraint motivates our research on geolocation, which is a topic that has received significant attention in the past decade.
However, most prior studies concentrate on user geolocation, which is estimating the home addresses of users\cite{qian2017probabilistic,tao2021episodic,zhou2022metageo,zhong2020interpreting}.
This type of geo-information is insufficient for applications like emergency location identification and natural disaster response\cite{li2018location}, which requires the location of individual posts.
Hence, in this paper, we focus on the problem of social post geolocation to infer the locations of individual posts.

For social post geolocation, previous efforts typically aim at inferring locations at the city level \cite{wing2014hierarchical,chi2016geolocation,li2018location}.
Although there is good performance at the city level, location information at such a coarse-grained level is still insufficient for the various applications mentioned earlier.
While some researchers studied the task of geo-coordinates estimation, it is challenging to achieve high accuracy\cite{mircea2020real,ouaret2019random}.
In real-life scenarios, semantic toponyms are more practical and understandable compared to numerical latitude and longitude\cite{wang2021semantic}.
Therefore, we study the problem of social post geolocation at the Point Of Interest (POI) level, a fine-grained semantic level.

However, Social Post Geolocation at POI level is a challenging problem due to the complexity, multi-modality, and diverse nature of social media data and their platforms.
Firstly, the user-generated textual content is short, free-form, and often noisy, containing acronyms, misspellings, and special tokens. 
It is non-trivial to understand such complex text precisely for location estimation.
Secondly, there are other non-textual contents such as time, social networks, images, and videos, which can be used for this task but also lead to the multi-modality issue. The ability to represent and fuse different data types is vital for geolocation.
Lastly, it is increasingly important to develop a geolocation framework with a generalization ability to deal with the emergence of diverse social platforms, like photo-sharing and micro-blogging platforms.
Many works focus on a single social platform with specific inputs, thus limiting their performance on other social platforms due to the difference in data fields. For better generalizability across platforms, some approaches utilize text content solely for geolocation but at the expense of missing out on other non-textual content and limiting performance.

To address these limitations, we present a transformer-based model, named transTagger, for POI-level social post geolocation, which is a general framework building upon Bidirectional Encoder Representations from Transformers (BERT) model with good generalization ability across different social platforms for accurate fine-grained location inference. The main contributions of this work can be summarized as follows:
\begin{itemize}
    \item We design a general categorization to tackle the multi-modality and diverse nature of social media data and their platforms, and provide four datasets with ground truth covering two cities and two platforms.
    \item We fuse features and learn their correlations using transformer encoders with a concatenated version of positional encodings, along with a novel temporal representation to provide an optimal combination strategy of representations for multi-modality fusion.
    \item We construct two variants, hierTagger and mtlTagger, by incorporating the hierarchy of locations into transTagger, and experimental results demonstrate that our models outperform state-of-the-art baselines by a considerate margin in terms of accuracy and distance error metrics.
\end{itemize}

The rest of the paper is organized as follows. 
In Section 2, we review the critical related work in the geolocation field and briefly introduce hierarchical classification techniques.
In Section 3, we first present the problem formulation and then describe our proposed model transTagger and two variants in detail.
Then Section 4 introduces the experimental setting, while Section 5 presents and discusses our experimental results.
Following that, we summarize and conclude this paper in Section 6.

\section{Related Work}
In this section, we review two main categories of work that are related to our research, namely social post geolocation and hierarchical geolocation works.

\subsection{Post Geolocation}
Post geolocation focuses on estimating the originating locations of social posts.
Unlike user geolocation, which leverages a user's entire posting history, post geolocation considers only an individual post or tweet and uses that as input.
For example, the work\cite{iso2017density} uses the convolutional mixture density network for location estimation with single tweet content.
Term co-occurrences in tweets, which exhibit spatial clustering or dispersion tendency, are detected and used to extend feature space in probabilistic language models\cite{ozdikis2018spatial}.
For location prediction during disaster events, Ouaret et al.\cite{ouaret2019random} present an iterative Random Forest fitting-prediction framework to learn semi-supervised models.
A name entity recognizer\cite{mircea2020real} is developed for geolocating tweets with the help of GeoNames gazetteer.
Kulkarni et al. \cite{kulkarni2020spatial} present a multi-level geocoding model that learns to associate texts with geographical locations and represent locations using S2 hierarchy.
Others propose to locate tweets based on BERT architecture with different tokenization settings, like vocabulary sizes\cite{scherrer2021social}.
In special cases, historical locations of users are involved to boost location inference performance, like using the Markov model to formalize tweet geolocation in a flood-related disaster based on history tweets\cite{singh2019event}.

Many researchers consider metadata to infer tweet locations\cite{chi2016geolocation,kordopatis2016placing}.
Pliakos and Kotropoulos construct a hypergraph based on images, users, geotags and tags of Flickr, which is further used for simultaneous image tagging and geolocation prediction\cite{pliakos2014simultaneous}.
A refined language model that is learned from massive corpora of social content, including tags, titles, descriptions, user ids, and image ids, is proposed to estimate the location of a post\cite{kordopatis2015geotagging}.
Miura et al.\cite{miura2016simple} propose a simple neural network structure with fully-connected layers and an average pooling process based on message text and user metadata for geolocation prediction.
To classify the microblogs of WeiBo into 8 semantic categories, the work\cite{wang2021semantic} explores the effect of user attributes and designs a neural network-based architecture with 4 feature fusion strategies.

\subsection{Hierarchical Geolocation}
Although the class hierarchy has been shown to be effective in closely relevant fields, like text classification\cite{zhou2020hierarchy,meng2019weakly,huang12019hierarchical,kowsari2017hdltex}, this problem has not so far received the attention it deserves.
Only a handful of existing works estimate locations of tweets and explore geolocation performance using hierarchical locations.
Previous efforts\cite{mahmud2012tweet,wing2014hierarchical} represent locations as a tree and construct a local classifier per parent node to infer locations, which corresponds to a typical hierarchical classification technique, Local Classifier per Parent Node (LCPN)\cite{silla2011survey}.
Multi-Task Learning (MTL) is incorporated to combine losses across multiple levels and predict locations at each level simultaneously\cite{huang2019hierarchical,kulkarni2020spatial}.
Most of these works aim at user location inference, whereas we study post geolocation.

Similar to our work, some research has tried to infer fine-grained locations of tweets\cite{chong2017exploiting,chong2018exploiting,mousset2020end}.
By investigating two properties, spatial focus and spatial homophily, a learning-to-rank framework\cite{chong2017exploiting,chong2018exploiting} is designed by ranking candidate venues.
The work\cite{mousset2020end} extracts semantic similarities between tweets and POI reviews locally and globally to provide a Spatially-aware Geotext Matching model building upon MLP.
Both methods have to compute similarity features explicitly with additional datasets, like check-in data or POI reviews from Foursquare, which is non-trivial and time-consuming.
While these works advance the task of tweet geolocation, our work differs from these earlier works in various ways, which we discuss next.
Our method takes as inputs tweet content and metadata of the Twitter dataset directly, building upon BERT and using transformer encoders to learn correlations among features.
Additionally, we employ a uniform representation of decomposed hierarchical time elements to further boost performance as the importance of temporal features is highlighted by many studies\cite{liu2018location,singh2019event,li2018location,mousset2020end}.
Moreover, we explore the effect of location hierarchy on the post geolocation performance by leveraging LCPN and MTL in our proposed models.

\begin{figure*}[!htbp]
  \centering
  \includegraphics[width=0.95\linewidth]{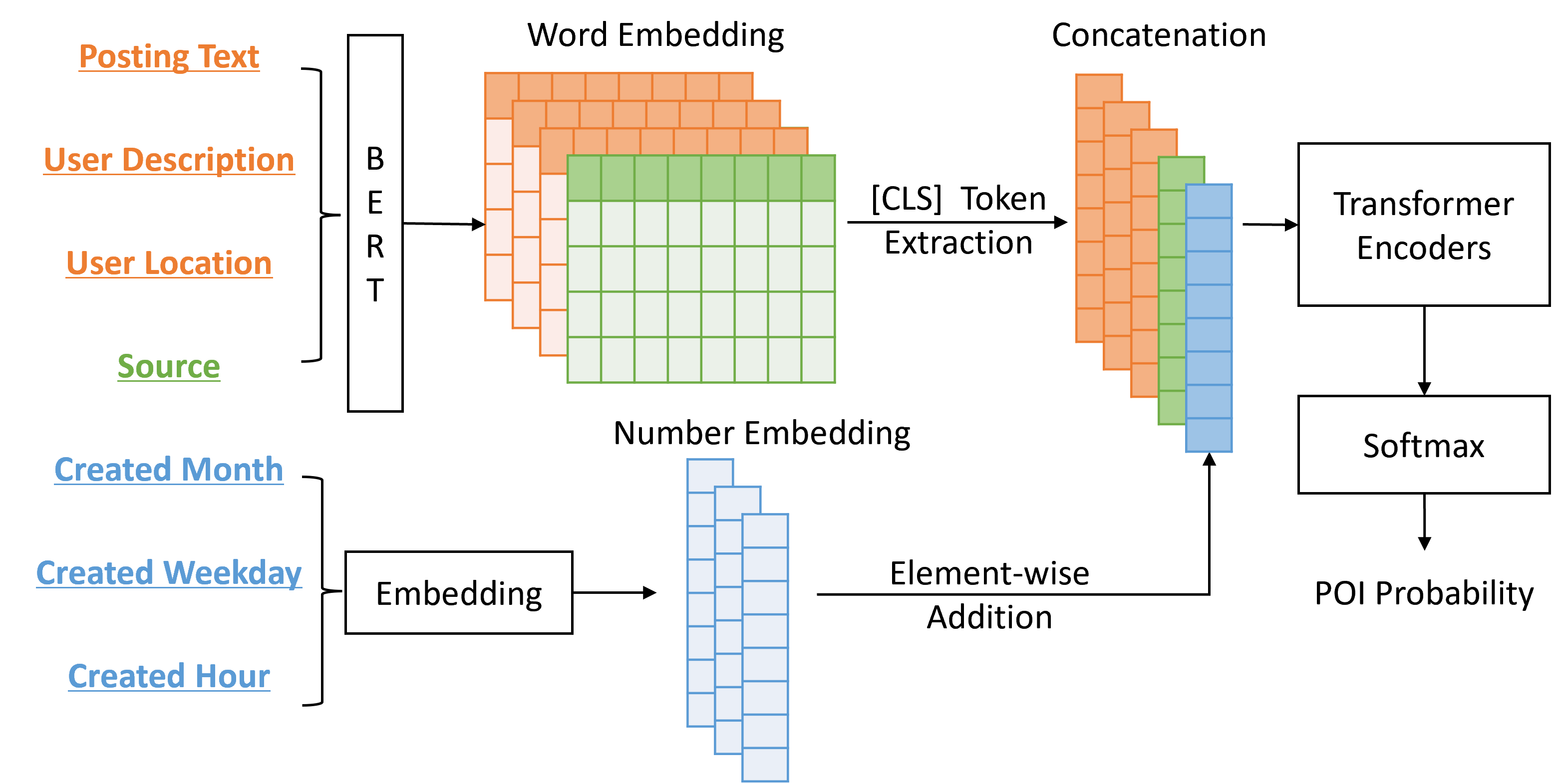}
  \caption{The architecture of our proposed model transTagger.}
  \label{transTagger}
\end{figure*}

\section{Method}
\subsection{Problem Formulation}
The \textbf{Social Post Geolocation} problem is defined as estimating the originating location of tweets. 
In the same spirit as prior studies\cite{li2018location,wing2014hierarchical,wang2021semantic}, the task is formulated as a classification problem where the predicted target is a location.
Unlike these earlier works, which classify posts into countries or cities, we aim at inferring locations at a finer-granularity level, that is at the landmark or POI level.
More specifically, the social post geolocation problem is represented as inferring POIs, given text and metadata of social media as input.


\subsection{Method Overview}
The overall structure of our proposed model, transTagger, is shown in Figure~\ref{transTagger}.
To tackle the inconsistency of different social platforms, we classify the inputs of social media data into three categories.
Information contained in social media data can be divided into user-generated and system-generated according to sources.
The user-generated content is of free form and could be very noisy. 
Besides posting texts, the user-generated content also includes user locations, user descriptions and so on. 
They form the first category of inputs and we denote it \textbf{Text}.
System-generated content comprises textual fields and numerical fields. 
The former is mostly categorical text, like source (indicating whether the tweet is posted from the phone or web platform), which falls into the second category of inputs: \textbf{Categorical Text (CT)}.
For the latter, numerical fields, a typical one is the time (when the post is created), and others are less explored and employed in post geolocation and we leave them for future research. 
The third category is \textbf{Time}, which we discuss the various representation techniques used later in this paper.
\textbf{Text}, \textbf{CT}, and \textbf{Time} are depicted in orange, green, and blue, respectively in Figure~\ref{transTagger}.

Our model applies BERT to learn semantic information and contextual information of \textbf{Text} and \textbf{CT}, and maps features into a word embedding space.
Following that, the representations of [CLS] tokens are extracted from all textual features and combined with embeddings of \textbf{Time}.
Then we use several layers of transformer encoders to learn the correlation of all features. 
The POI probability of each post is calculated using a fully-connected layer with the activation function as softmax.
We introduce the detailed implementation of our model transTagger in the following sections.

\subsection{Feature Representation}
We apply the pre-trained model by plugging in the post geolocation task-specific inputs and outputs into BERT. 
At the architecture level, BERT is an $L$-layer bi-directional transformer encoder\cite{devlin2018bert}.
The hidden size and the number of self-attention heads for each component are denoted as $H$ and $A$, respectively.

\textbf{Text}, including posting texts, user locations, user descriptions, and \textbf{CT}, like sources, are all used as inputs.
Here a degenerate text-$\varnothing$ pair corresponds to sentence $A$ and sentence $B$ since we formulate the post geolocation task as a classification problem and there is no "sentence" pair.
An input sample is regarded as a sentence in this paper although it may actually contain multiple sentences.
During tokenization, each sentence is converted into a sequence of tokens and a special classification token, [CLS], is injected in front of every input sample\cite{devlin2018bert}.
Then the first token becomes [CLS].
Besides the above token embedding, other embeddings are utilized to take the position information inside sentences or between sentence pairs into consideration.
Position embedding represents the position of each token in a sentence.
In contrast, segment embedding is used to distinguish sentences $A$ and $B$ and thus is set to all zero in our case.
The element-wise addition of token, position and segment embeddings forms the input representation\cite{devlin2018bert}.

We denote the learned embedding in the final hidden layer of each input sample as $E\in\mathbb{R}^{N\times H}$ where $N$ is the sentence length.
The corresponding embedding of the [CLS] token is represented as $C\in\mathbb{R}^{H}$.
This token embedding can be seen as the aggregation of sentence representation, which is used for subsequent application.
Note that all the parameters are fine-tuned in an end-to-end manner based on our task, post geolocation.

Time is a vital factor in relation to human mobility and, thus, of great importance for location inference. 
However, most works simply represent it as one-hot encoding based on the timestamp, which does not capture the full extent of temporal information and ignores the hierarchy of time elements, like hours and months.
Inspired by this work\cite{zhou2021informer}, we propose a uniform representation of hierarchical time elements, UniHier, to learn temporal information. 
Hierarchical time elements are extracted from \textbf{Time}, including hours, weekdays, and months.
Then each element is represented as a learnable embedding vector with dimension $H$ and limited vocab size. 
A uniform representation of time is constructed by the element-wise addition of all embedding vectors.

\subsection{Feature Fusion}
Assuming that \textbf{Text} contains $m$ fields, \textbf{CT} contains $n$ fields, we extract [CLS] token vectors of \textbf{Text} and \textbf{CT}, and concatenate them with the UniHier representation of \textbf{Time}, then a feature matrix $F\in\mathbb{R}^{(m+n+1)\times H}$ is generated.

To learn the correlation of all features, we employ a multi-layer transformer encoder as described in the work\cite{vaswani2017attention}.
Positional encodings are represented using sine and cosine functions of different frequencies as below and $pos$ is the position, $i$ is the dimension:
\begin{align}
    PE_{(pos, 2i)} &= sin(pos/10000^{2i/H}) \\
    PE_{(pos, 2i+1)} &= cos(pos/10000^{2i/H})
\end{align}
These positional encodings are fixed during training and with dimension $H$.
Different from the now ubiquitous transformer encoder, which sums feature representations and the corresponding positional encodings, we concatenate them and call it the concatenated version of positional encodings.
Experiments demonstrate that this approach improves performance.

After concatenating with positional encodings, this feature matrix is utilized to calculate POI probabilities with a softmax layer. This model is then trained using the Adam update rule as the optimizer.

\subsection{Hierarchical Prediction}
The hierarchy of locations enables the application of hierarchical prediction and thus improves the performance of post geolocation.
We incorporate LCPN, a typical hierarchical classification approach, with transTagger, and construct a variant, hierTagger.
By combining the class hierarchy with MTL, we build upon our earlier described transTagger and propose another variant, mtlTagger.
Next, we elaborate on the details behind hierTagger and mtlTagger.

\subsubsection{hierTagger}
The LCPN approach aims to train a multi-class classifier for each parent node in the class hierarchy, to distinguish between its child nodes\cite{silla2011survey}.
The class hierarchy is typically a tree or a Direct Acyclic Graph (DAG), which is represented as a tree in our case.
We build the tree of toponyms at different scales, from coarse to fine, starting from a root node that covers the whole research area.
For every parent node in this tree, we employ transTagger to construct a local classifier, which is trained independently.
Then a top-down class prediction approach is applied during the testing phase.

\subsubsection{mtlTagger}
MTL provides models with better generalization ability by sharing representations between related tasks\cite{ruder2017overview}. 
Tasks predicting posts' locations at coarser levels are designed as auxiliary tasks.
We incorporate transTagger with hard parameter sharing, the most commonly used approach to MTL in neural networks, to predict posts' locations at different scales, from coarse to fine.
The prediction result for the coarser level, denoted as $q$, is further utilized to constrain the finer level prediction by adding $q$ to the loss function of the finer level.
A correlation matrix between the two levels is employed to help the loss function of the finer level understand the coarser level's prediction result. 

\section{Experimental Setting}
\subsection{Datasets}
We perform our experiments using datasets from two different social media platforms, Flickr and Twitter, for two cities of Melbourne and Singapore.

\subsubsection{Twitter}
We collected 266,614 geotagged tweets that were posted in Melbourne from 2010 to 2018, and 482,765 geotagged tweets that were posted in Singapore from 2018 to 2022. 
We also combined tweets from Melbourne and Singapore for experiments to test the robustness of our models.
The Twitter datasets of Melbourne, and Singapore, and their combination are denoted as Twitter-Mel, Twitter-SG, and Twitter-SM, respectively.

\subsubsection{Flickr}
The Flickr dataset comprises 78,131 geotagged images that were posted in Melbourne from 2004 to 2020, extracted using the Flickr API and from the Yahoo! Flickr Creative Commons 100M (YFCC-100M)\footnote{http://research.yahoo.com/Academic\_Relations}. We further augmented this dataset by collecting the metadata of Flickr users. This dataset is denoted as Flickr-Mel.\footnote{We also collected a Flickr dataset for Singapore but excluded it for further experimentation due to a low number of data points.}

A list of POIs and their categories are obtained using the Google Place API\footnote{https://developers.google.com/maps/documentation/places/web-service/overview}.
For Singapore, our research area is the whole country/city and there are 9,666 POIs.
For Melbourne, we concentrate on the central city area and there are 242 POIs.
To implement hierarchical prediction, POI themes and POI sub-themes are involved as labels to construct the class hierarchy.
Specifically, there are 16 POI themes (eg., Leisure/Recreation), 49 POI sub-themes (eg., Park/Garden), and 242 POIs (eg., Batman Park). 

Our work aims to predict the posts' POIs within a specific city, Melbourne and Singapore in our case, in contrast to existing efforts that focus on coarse-level predictions, at the city, country, or even continent level.
To this end, we label a tweet $tw$ in the Twitter dataset (or image $im$ in the case of the Flickr dataset) as one and only one POI.
Following the proximity principle\cite{taylor2018travel}, we compare the distance between $tw$ (or $im$) and the POI location using their latitude and longitude coordinates, and label it with the POI if their distance differs by less than 100 meters.
Any $tw$ (or $im$) that is not assigned a POI label is then filtered out.
Note that the above statistics of the Twitter and Flickr datasets are computed after POI-labelling preprocessing.

Our two variants involve the use of class hierarchy of POIs. For example, hierTagger utilizes POI-theme level and POI-level labels, while mtlTagger contains three loss functions that are designed for POI theme, POI sub-theme, and POI predictions, respectively. 

\subsection{Evaluation Metrics}
We use two evaluation metrics that are frequently used in geolocation tasks, namely accuracy and distance error.
Accuracy, denoted as $\boldsymbol{acc@k}$, reflects the proportion of correct predictions based on the top-k results and we evaluate with $k$ as 1, 5, 10, and 20.
Mean distance error, represented as $\boldsymbol{mean}$, measures the mean distance between the predicted location and actual POI location.
We also experimented using median distance error and observe that our models achieve 0 error, thus we do not report the results for concision.
To make our models comparable to the baselines, we evaluate metrics at the POI level although there are predictions at other scales (which the baseliens are unable to account for).

\subsection{Parameter Setting}
In our experiments, the max sequence length for text and other textual features is 100.
To represent \textbf{Time} inputs using UniHier, they are randomly initialized from a uniform distribution $U(-1.0, 1.0)$ with dimension 128 (this value corresponds to the dimension of word embeddings) and vocab size is limited to 60 since the finest granularity is a minute.
These embeddings are then learned during training.

The hyperparameter tuning is conducted using Bayesian optimization on the learning rate, the number of encoder layers, the number of heads, hidden size, and batch size.
The number of layers, the number of attention heads, and the hidden size of the transformer encoder before the softmax layer are set as 3, 48, and 1300, respectively.
The training of our model is performed using Adam with an initial learning rate of 3e-4 and a batch size of 128.
We train the model with 4 epochs.
Additionally, the block threshold for hierTagger is set as 0.01 and the loss weights for mtlTagger are 0.1, 0.1, and 1.

\subsection{Baselines}
We compare our proposed model and two variants with various popular geolocation models, including \textbf{MNB-Ngrams} (Multinomial Naive Bayes with Uni/Bi/Tri-grams) \cite{chi2016geolocation,chong2018exploiting,han2014text,mahmud2012tweet,ozdikis2018spatial}, \textbf{CNN-TT} (Convolutional Neural Network with Text and Time) \cite{lim2019geotagging}, and \textbf{HLPNN} (Hierarchical Location Prediction Neural Network)\cite{huang2019hierarchical}. 
The CNN text classification model \cite{kim-2014-convolutional} is widely used for geolocation\cite{huang2017predicting,liu2018location,iso2017density} and thus we include it and one of its variants into our experiments as \textbf{CNN} and \textbf{CNN-1Hot} \cite{Johnson2015EffectiveUO}, respectively.
Besides HLPNN, one more hierarchical classification model, \textbf{HDLTex} (Hierarchical Deep Learning for Text Classification)\cite{kowsari2017hdltex} is utilized as one of baselines. 
Our two proposed variants, hierTagger and mtlTagger, are also involved in comparisons.

\begin{table*}[!htbp]
\centering
\caption{Baseline comparison on Flickr-Mel and Twitter-Mel.}\label{baselines1}
\resizebox{1\textwidth}{!}{
\begin{tabular}{lllllllllll} 
\hline 
&\multicolumn{5}{c}{Flickr-Mel}&\multicolumn{5}{c}{Twitter-Mel}\\
 &Acc@1$\uparrow$ &Acc@5$\uparrow$ &Acc@10$\uparrow$ &Acc@20$\uparrow$ &Mean(m)$\downarrow$ &Acc@1$\uparrow$ &Acc@5$\uparrow$ &Acc@10$\uparrow$ &Acc@20$\uparrow$ &Mean(m)$\downarrow$\\
\hline
HLPNN&	68.68&	83.62&	88.95&	93.87&	247.6&	61.45&	76.7&	81.85&	87.05&	433.2\\
HDLTex&	56.89&	64.71&	66.49&	70.14&	604&	56.2&	64.67&	66.33&	67.69&	512.5\\
CNN-TT&	75.49&	87.63&	90.83&	94.14&	241&	67.85&	80.69&	84.93&	89.19&	351.9\\
CNN&	59.4&	74.19&	81.16&	88.43&	528&	60.45&	77.27&	83.45&	88.54&	408.5\\
CNN-1Hot&	59.91&	76.69&	83.25&	90.14&	697.7&	63.08&	76.89&	80.92&	85.43&	362\\
MNB-Ngrams&	54.35&	71.71&	79.93&	88.61&	1071&	49.82&	73.6&	79.05&	84.62&	500.7\\
\hline
transTagger&	\textbf{77.88}&	89.85&	\textbf{93.05}&	93.05&	\textbf{175.8}&	\textbf{71.96}&	84.64&	\textbf{88.2}&	88.2&	\textbf{303.3}\\
hierTagger&	77.59&	\textbf{90.13}&	92.91&	\textbf{95.87}&	183.5&	71.84&	\textbf{84.67}&	88.03&	91.44&	317.9\\
mtlTagger&	77.22&	89.44&	92.86&	95.73&	182.9&	71.42&	84.34&	88.12&	\textbf{91.49}&	319.5\\
\hline
\end{tabular}
}
\end{table*}

\begin{table*}[!htbp]
\centering
\caption{Baseline comparison on Twitter-SG and Twitter-SM.}\label{baselines2}
\resizebox{1\textwidth}{!}{
\begin{tabular}{lllllllllll} 
\hline 
&\multicolumn{5}{c}{Twitter-SG}&\multicolumn{5}{c}{Twitter-SM}\\
 &Acc@1$\uparrow$ &Acc@5$\uparrow$ &Acc@10$\uparrow$ &Acc@20$\uparrow$ &Mean(km)$\downarrow$ &Acc@1$\uparrow$ &Acc@5$\uparrow$ &Acc@10$\uparrow$ &Acc@20$\uparrow$ &Mean(km)$\downarrow$\\
\hline
CNN-TT&	53.76&	67.27&	70.29&	72.81&	2.617&	54.8&	68.64&	72.41&	75.7&	154.3\\
CNN&	49.77&	62.04&	64.72&	67.66&	2.949&	48.98&	62.11&	65.5&	69.09&	542.7\\
CNN-1Hot&	38.34&	50.11&	52.96&	55.86&	3.536&	38.85&	52.05&	55.54&	59.21&	882.9\\
\hline
transTagger&	\textbf{61.94}&	\textbf{73.75}&	\textbf{76.75}&	\textbf{76.75}&	\textbf{2.215}&	\textbf{64.88}&	\textbf{76.8}&	\textbf{80.08}&	\textbf{80.08}&	\textbf{3.69}\\
\hline
\end{tabular}
}
\end{table*}

\section{Experimental Results}
\subsection{Baseline Comparison}
To verify the effectiveness of our proposed models, experiments are designed to compare the performance of three variants and various baselines on datasets Flickr-Mel and Twitter-Mel, as shown in Table~\ref{baselines1}.
Similar experiments are conducted on datasets Twitter-SG and Twitter-SM as well to further examine the robustness of geolocation performance as presented in Table~\ref{baselines2}.
We only report results for transTagger and three baselines among all variants and baselines as the hierarchical labels are not available for these two datasets.

Overall, transTagger, hierTagger, and mtlTagger outperform all baselines, including the hierarchical ones, across four datasets.
Compared with a strong baseline like CNN-TT, transTagger outperforms by a substantial margin, obtaining 2.39\%, 4.11\%, 8.18\%, and 10.08\%, improvement in accuracy (acc@1) on Flickr-Mel, Twitter-Mel, Twitter-SG, and Twitter-SM.
The latter two datasets contain many more POIs and the improvement of transTagger over the baselines is even larger.
This indicates that our model is versatile to handle a large number of classes (POIs) well.
In addition to accuracy, the mean distance error is also reduced greatly.
To be specific, transTagger reduces the mean distance error by 65.2, 48.6, 402, and 1174 meters, compared with CNN-TT.
Notably, values of mean distance error on Twitter-SG and Twitter-SM are relatively higher compared with Flickr-Mel and Twitter-Mel because they cover much larger areas.
Thus we use different units, meters (m) and kilometers (km) in  Table~\ref{baselines1} and Table~\ref{baselines2} to denote distance.

As shown by the results, our proposed models provide superior performance for POI-level post geolocation across all cities and platforms, compared to the various baselines.

\begin{table*}[!htbp]
\centering
\caption{Representation combination selection on Flickr-Mel and Twitter-Mel.}\label{rcs1}
\resizebox{1\textwidth}{!}{
\begin{tabular}{lllllllllll} 
\hline 
&\multicolumn{5}{c}{Flickr-Mel}&\multicolumn{5}{c}{Twitter-Mel}\\
 &Acc@1$\uparrow$ &Acc@5$\uparrow$ &Acc@10$\uparrow$ &Acc@20$\uparrow$ &Mean(m)$\downarrow$ &Acc@1$\uparrow$ &Acc@5$\uparrow$ &Acc@10$\uparrow$ &Acc@20$\uparrow$ &Mean(m)$\downarrow$\\
\hline
\textit{transTagger}\\
Text-Text&	77.88&	89.85&	93.05&	93.05&	175.8&	\textbf{71.96}&	\textbf{84.64}&	88.2&	88.2&	\textbf{303.3}\\
1Hot-Text&	77.88&	89.85&	93.05&	93.05&	175.8&	71.69&	84.6&	\textbf{88.29}&	\textbf{88.29}&	322.1\\
Text-UniHier&	\textbf{78.04}&	\textbf{90.16}&	\textbf{93.28}&	\textbf{93.28}&	\textbf{171.6}&	70.03&	84.32&	88.09&	88.09&	313.8\\
1Hot-UniHier&	\textbf{78.04}&	\textbf{90.16}&	\textbf{93.28}&	\textbf{93.28}&	\textbf{171.6}&	69.69&	84.25&	87.87&	87.87&	308.3\\
Text-1Hot&	77.49&	89.66&	92.99&	92.99&	184.1&	69.91&	84.13&	87.71&	87.71&	316.9\\
1Hot-1Hot&	77.49&	89.66&	92.99&	92.99&	184.1&	69.5&	84.26&	88.04&	88.04&	321.6\\
\hline
\textit{hierTagger}\\
Text-Text&	77.59&	\textbf{90.13}&	92.91&	\textbf{95.87}&	183.5&	71.42&	84.34&	88.12&	91.49&	319.5\\
1Hot-Text&	77.59&	\textbf{90.13}&	92.91&	\textbf{95.87}&	183.5&	\textbf{71.49}&	\textbf{84.45}&	\textbf{88.15}&	\textbf{91.56}&	324.5\\
Text-UniHier&	\textbf{78.18}&	89.94&	\textbf{93.15}&	95.83&	\textbf{169.5}&	70.03&	84.29&	88.03&	91.52&	314.4\\
1Hot-UniHier&	\textbf{78.18}&	89.94&	\textbf{93.15}&	95.83&	\textbf{169.5}&	69.6&	84.18&	87.78&	91.07&	\textbf{308.8}\\
Text-1Hot&	77.23&	89.43&	92.82&	95.56&	190.2&	69.82&	84.04&	87.57&	91.16&	316.7\\
1Hot-1Hot&	77.23&	89.43&	92.82&	95.56&	190.2&	69.35&	84.19&	87.96&	91.46&	321.9\\
\hline
\textit{mtlTagger}\\
Text-Text&	77.22&	89.44&	92.86&	95.73&	182.9&	\textbf{71.84}&	\textbf{84.67}&	88.03&	91.44&	317.9\\
1Hot-Text&	77.22&	89.44&	92.86&	95.73&	182.9&	71.48&	84.45&	\textbf{88.04}&	\textbf{91.64}&	315.1\\
Text-UniHier&	\textbf{78.93}&	\textbf{90.18}&	93.31&	95.97&	\textbf{168.3}&	69.91&	84.3&	87.94&	91.52&	\textbf{312.9}\\
1Hot-UniHier&	\textbf{78.93}&	\textbf{90.18}&	93.31&	95.97&	\textbf{168.3}&	69.16&	84.06&	88.01&	91.39&	314.2\\
Text-1Hot&	77.84&	89.9&	\textbf{93.36}&	\textbf{96.26}&	179.1&	69.62&	84.18&	87.83&	91.35&	317.3\\
1Hot-1Hot&	77.84&	89.9&	\textbf{93.36}&	\textbf{96.26}&	179.1&	69.39&	84&	87.72&	91.33&	314.9\\
\hline
\end{tabular}
}
\end{table*}

\begin{table*}[!b]
\centering
\caption{Representation combination selection of transTagger on Twitter-SG and Twitter-SM.}\label{rcs2}
\resizebox{1\textwidth}{!}{
\begin{tabular}{lllllllllll} 
\hline 
&\multicolumn{5}{c}{Twitter-SG}&\multicolumn{5}{c}{Twitter-SM}\\
 &Acc@1$\uparrow$ &Acc@5$\uparrow$ &Acc@10$\uparrow$ &Acc@20$\uparrow$ &Mean(km)$\downarrow$ &Acc@1$\uparrow$ &Acc@5$\uparrow$ &Acc@10$\uparrow$ &Acc@20$\uparrow$ &Mean(km)$\downarrow$\\
\hline
Text-Text&	\textbf{61.94}&	\textbf{73.75}&	\textbf{76.75}&	\textbf{76.75}&	\textbf{2.215}&	\textbf{64.88}&	76.8&	\textbf{80.08}&	\textbf{80.08}&	\textbf{3.69}\\
1Hot-Text&	61.37&	73.26&	76.36&	76.36&	2.292&	64.84&	\textbf{76.88}&	80.06&	80.06&	3.263\\
Text-UniHier&	58.1&	72.71&	75.92&	75.92&	2.318&	61.9&	76.1&	79.55&	79.55&	56.63\\
1Hot-UniHier&	57.82&	72.63&	75.94&	75.94&	2.332&	61.53&	76.13&	79.48&	79.48&	69.64\\
Text-1Hot&	58.13&	72.71&	75.92&	75.92&	2.334&	61.74&	76&	79.33&	79.33&	67.52\\
1Hot-1Hot&	57.3&	72.43&	75.65&	75.65&	2.349&	61.21&	75.8&	79.24&	79.24&	58.33\\
\hline
\end{tabular}
}
\end{table*}


\subsection{Representation Combination Selection}
Taking the generalization into consideration, we categorize inputs into three types: \textbf{Text}, \textbf{CT}, and \textbf{Time}, and describe how to do representation for each type in Method section.
However, there could be multiple ways to represent each type.
For \textbf{CT}, one way is to treat categorical texts as normal texts and use BERT or other language models to generate representations, as described in the Method section, and we call this Text embedding.
Another commonly used way is one-hot encoding.
For \textbf{Time}, one way is to treat date time as a standard text and generate temporal embedding using language models.
Hence, there are two ways to represent \textbf{CT}: text and one-hot,  and three ones for \textbf{Time}: text, one-hot, and UniHier, and we experiment with all six combinations of these representation methods to find an optimal representation combination strategy. 
The results are illustrated in Table~\ref{rcs1} and Table~\ref{rcs2}, where Text denotes Text embedding, and Text-UniHier means doing Text embedding for \textbf{CT} and using UniHier representation for \textbf{Time}, and so forth.
Note that the results of Text-Text and 1Hot-Text are duplicated for Flickr since there are no \textbf{CT} fields. So do Text-UniHier and 1Hot-UniHier, Text-1Hot and 1Hot-1Hot.

Text-Text delivers the overall best performance across all Twitter datasets. 
But for the Flickr dataset, Text-UniHier (or 1Hot-UniHier) outperforms others.
One possible reason is that Flickr contains more time fields, including photo taken time and photo posted time, than Twitter, with only tweet created time.
Thereby, the best representation combination is Text-Text. 
In case of involving in many time inputs, it is recommended to represent temporal inputs using UniHier.

Comparing the three variants, hierTagger and mtlTagger show no advantage except for acc@20.
Hence, these two variants are recommended when this metric matters a lot.


\begin{table*}[!htbp]
\centering
\caption{Ablation study on Twitter-SG and Twitter-SM.}\label{ablation2}
\resizebox{1\textwidth}{!}{
\begin{tabular}{lllllllllll} 
\hline 
&\multicolumn{5}{c}{Twitter-SG}&\multicolumn{5}{c}{Twitter-SM}\\
 &Acc@1$\uparrow$ &Acc@5$\uparrow$ &Acc@10$\uparrow$ &Acc@20$\uparrow$ &Mean(km)$\downarrow$ &Acc@1$\uparrow$ &Acc@5$\uparrow$ &Acc@10$\uparrow$ &Acc@20$\uparrow$ &Mean(km)$\downarrow$\\
\hline
transTagger&	61.94&	73.75&	76.75&	76.75&	2.215&	64.88&	76.8&	80.08&	80.08&	3.69\\
w/o transformer&	60.3&	72.44&	75.64&	75.64&	2.338&	63.92&	76.7&	80.1&	80.1&	6.108\\
w/o position&	61.28&	73.01&	75.96&	75.96&	2.292&	64.39&	76.43&	79.75&	79.75&	4.917\\
\hline
\end{tabular}
}
\end{table*}

\subsection{Ablation Study}
We compare transTagger with two ablations to examine the effectiveness of two model components, namely transformer encoders and position encodings.
Table~\ref{ablation2} shows the performance breakdown on Twitter-SG and Twitter-SM.
For w/o position, we replace the concatenation version of positional encodings with the commonly used add-on version.
The w/o transformer ablation removes the transformer encoders which are used to learn the correlation of features.
The results demonstrate that all components contribute to improving the post geolocation performance of transTagger.
Out of all these, encoders have the greatest effect as shown by how it increases accuracy (including acc@1, acc@5, acc@10, and acc@20) and reduces the mean distance error by the largest margin.

\section{Conclusion}
In this paper, we propose a transformer-based general framework, transTagger, for POI-level post geolocation.
The inputs are categorized into three types: \textbf{Text}, \textbf{CT}, and \textbf{Time} to handle different social data, and the optimal representation combination, Text-Text, is provided by experimenting with all combinations.
A novel representation of time, UniHier, is presented and verified to be useful in the case of many temporal inputs.
Transformer encoders are employed to enhance geolocation performance and a concatenated version of encodings is incorporated to capture feature-wise positions.
The effectiveness and robustness of our model are demonstrated on four datasets, covering two cities and two social platforms.
Two variants, hierTagger and mtlTagger, by incorporating respective LCPN and MTL with transTagger, are shown to lift acc@20 effectively.

\section{Acknowledgments}

This research is funded in part by the Singapore University of Technology and Design under grant SRG-ISTD-2018-140.
The computational work was partially performed on resources of the National Supercomputing Centre, Singapore.


\bibliographystyle{splncs04}
\bibliography{transBERT}

\end{document}